# A Comparative Performance Study of the Routing Protocols RPL, LOADng and LOADng-CTP with Bidirectional Traffic for AMI Scenario

Saida Elyengui, Riadh Bouhouchi, and Tahar Ezzedine

***Abstract*—With the introduction of the smart grid, Advanced Metering Infrastructure (AMI) has become a main component in the present power system. The effective implementation of AMI depends widely on its communication infrastructure and protocols providing trustworthy two-way communications. In this paper we study two routing protocols philosophies for low power and lossy networks (LLNs) and their application for a smart metering scenario. This study purposes a detailed evaluation of two routing protocols proposed by IETF, the proactive candidate namely RPL (IPv6 Routing Protocol for Low-Power and Lossy Networks) and the reactive candidate named LOADng (LLN On-demand Ad-hoc Distance vector routing protocol – next generation) recently proposed as an internet Draf , still in its design phase and is part of the ITU-T G.9903 recommendation. In the course of this study, we also implemented an extension version of LOADng, named LOADng-CTP specified by an IETF draft extended with a collection tree for efficient data acquisition in LLNs. We performed checks on control overhead; End to End Delay and Packet delivery ratio for the two protocols related to multipoint-to-point (MP2P), and point-to-multi point (P2MP) traffic flow in a realistic smart metering architecture.**

***Index Terms*—Smart Metering; RPL; LOADng; LOADng-CTP; Performance; Simulation; Contiki OS; Cooja.**

## I. Introduction

In the recent last year's we were witnessing rapid transformation in worldwide electric grid to meet new requirements such as efficiency, reliability, increasing power demand and sustainability [1, 2]. However, in order to address the different challenges for these new requirements, a novel distributed bidirectional model is mandatory for electricity generation, production distribution and control. Indeed, utility companies need to deploy a robust two-way flow communication infrastructure between electricity consumers and providers and in order to meet the new requirements for different Smart grid applications like Advanced Metering Infrastructure (AMI) and smart homes and buildings [3]. Networks connecting smart devices in Low Power and Lossy Networks (LLNs), such as sensors and actuators, operate frequently in extremely capricious link quality conditions and very restricted internal (memory and CPU) and limited communications capacity and constrained energy resources. So far, routing protocols are required for creating and sustaining multi-hop connectivity in LLNs with respect to QoS requirements for such applications.

In this manuscript we make the following contributions: We evaluate our implementations of LOADng and LOADng-CTP in Contiki OS and compare it to RPL protocol for bidirectional scenarios in AMI mesh network architecture. We provide simulation results for the network End-to-End delay, PDR, overhead and show how our implemented LOADng-CTP solution can provide bidirectional data flow scalability for AMI scenario.

The paper is structured as follows: Section II depicts related work performed by the community. Section III provides a detailed overview of RPL protocol. Respectively in sections IV and V we describe LOADng and LOADN-CTP protocols specifications. Section VI presents detailed performance evaluation and simulations results, in section VII we conclude.

## II. Related Work

The authors in [4] proposed an extension to LOADng routing protocol named LOADng Collection Tree and performed a detailed comparison of RPL, LOADng and LOADng-Collection Tree in terms of overhead, Packet delivery ratio and End to End Delay. In [5] the authors performed a detailed study of proactive versus reactive routing in low power and lossy Networks and depicted performance analysis and scalability improvements for RPL and LOADng protocols. In their study of the two protocols philosophies several metrics of interest were investigated with varied size deployments scenarios and traffic flow, as well as random topologies. They concentrated on metrics related to scalability such as End to End delay, Hop Distance and Control overhead. In [6] the authors proposed the LOADng-CTP protocol the Collection Tree Extension of Reactive Routing Protocol for Low-Power and Lossy Networks. The protocol complexity, interoperability and security were deeply examined. The comparative study was performed with respect to metrics such as packet delivery ratio, Number of MAC layer collisions, Average End-to-End delay, Number of overhead packets transmitted by each router and Overhead bytes per second in the whole network. The authors proved that LOADng-CTP extension can efficiently increase the proficiency of data acquisition in LLNs networks.

Authors are with Communication System Laboratory Sys'Com, National Engineering School Of Tunis, Tunis El Manar University, Tunisia (e-mail: saida.yengui@gmail.com;Riadh.bouhouchi@yahoo.com;Tahar.Ezzedine@enit.rnu.tn).

## III. RPL Specification

The Routing protocol for Low power and Lossy networks RPL was developed and published as (RFC6550) [7] by the ROLL Working Group of the IETF. In essence, RPL is a proactive distance-vector routing protocol for IPv6-based LLNs. RPL constructs its routes in periodic intervals and forms a tree-like routing structure in the network called Destination Oriented Directed Acyclic Graph (DODAG) defined according to a specific objective function and a set of metrics and constraints. A network can be established by multiple DODAG. A Directed Acyclic Graph (DAG) is formed over a mesh network with sink nodes that serve as roots of the graph. RPL divides the DAG into one or more DODAGs and attributes only one sink per DODAG. The protocol operation mode can be divided into two main phases; Routing upward or DODAG construction, and Routing downward/Destination Advertisement. In the first step, RPL provisions routes Up towards DODAG roots; each node chooses its preferred parent in the tree to convey DODAG information, it uses DODAG Information Object (DIO) and DODAG Information Solicitation (DIS) messages [2]. Then, RPL uses Destination Advertisement Option (DAO) messages to establish downward routes. In the DODAG construction process, topology building is managed by several rules such as Loop avoidance, a chosen Operating function OF, path metrics etc. First of all, upon the trickle timer expiration a DIO packet is sent by the RPL node in local multicast, in order to advertise a DODAG end its characteristics like DODAGID and DODAG Rank of the node. When a node receives a DIO packet, it must verify the message. If the packet is considered for processing, the RPL node determines whether it is send from a candidate neighbor or not. Then, it checks whether the packet is related to a DODAG that the current node is already member of. After checking the rank of the current link, the packet is considered processed. Then, in terms of Destination Advertisement there are two modes for downward traffic: Storing when the packets have to travel all the way to a DODAG root before traveling down, or Non-Storing, in this case, the packet may be directed down towards the destination by a common ancestor of the source. RPL supports three basic traffic flows Multipoint-to-Point Traffic (M2P), Point-to-Multipoint Traffic (P2M), Point-to-Point Traffic (P2P). All of them are definitely required in several applications like smart grid applications, building and home automation [8], and smart metering. As upwards roots are stored in routing tables, it is well optimized for M2P. Yet, it gives -by non-storing mode - reasonably support for P2MP, besides, it provides only basic features for P2P [9].

## IV. LOADng Specification

LLNs Ad hoc On-Demand Distance Vector Routing – Next Generation LOADng is a reactive distance-vector routing protocol for WSNs [10]. It is a simplified version of ad-hoc on-demand routing protocol AODV originally developed for use in IEEE 802.15.4 based devices in 6LoWPANs and LLNs. This protocol may be used at layer 3 as a route-over routing protocol or at layer 2 as a mesh-under protocol. Therefore, LOADng algorithm is characterized by its simplicity and its low memory storage needs. Thus, it would be ideal and suitable solution for AMI mesh networks. As it was originally developed to WSNs and Low Power and Lossy Networks (LLNs), it should be adapted to their requirements and constraints.

LOADng describes four types of packets:

- Route Request (RREQ): The RREQ is generated by a router the <originator>, when a data packet in available to a destination, RREQ packet is with no valid route and with a specific destination address.
- Route Reply (RREP): The RREP is generated by a router, upon a RREQ reception and processing with destination address in its routing set.
- Route Reply Acknowledgement (RREP-ACK): The RREP-ACK is generated by a LOADng router after a reception of RREP, as an indication to the neighbor source of the RREP that the RREP was successfully received.
- Route Error (RERR): the RERR is generated by a router when the router detects a broken route to the destination.

LOADng inherited basic operations of AODV, including generation and forwarding of Route Request RREQs as described in Fig. 1 to discover a route to a specific destination.

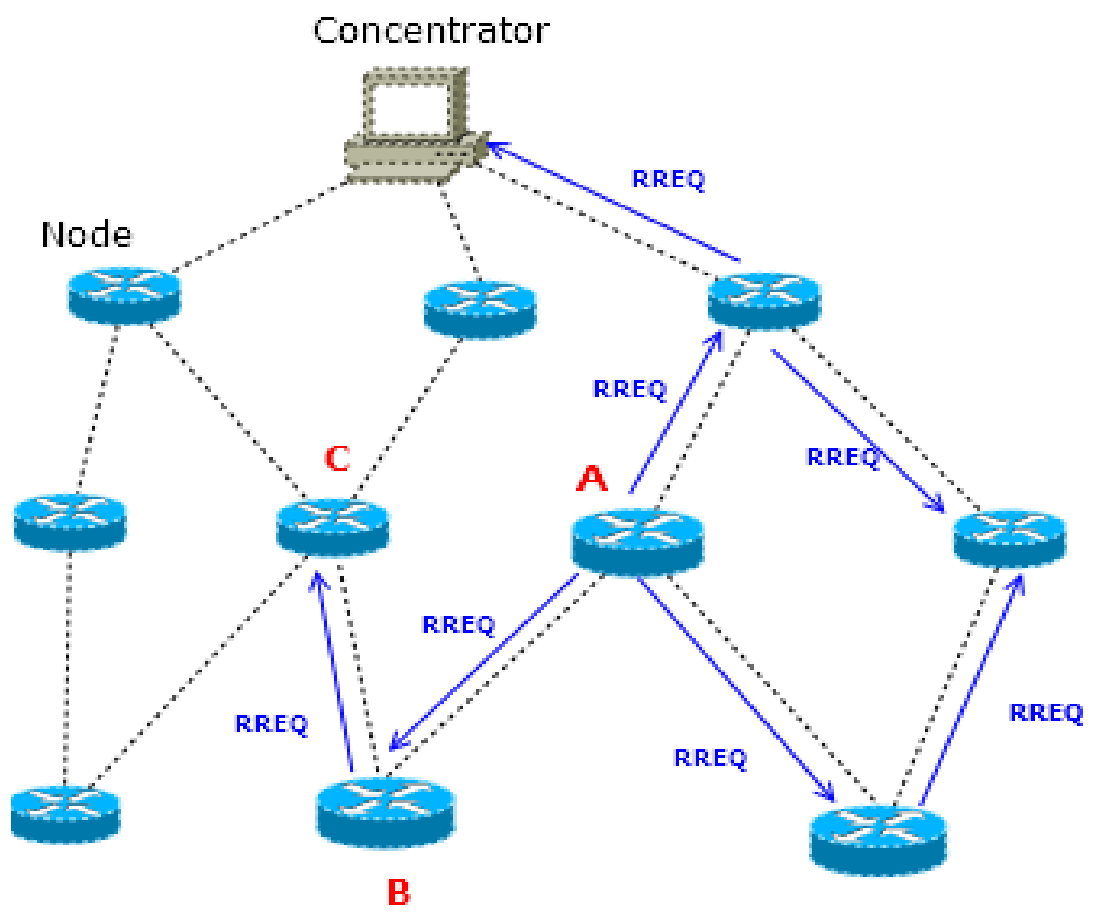

Fig. 1. RREQ forwarding in LOADng

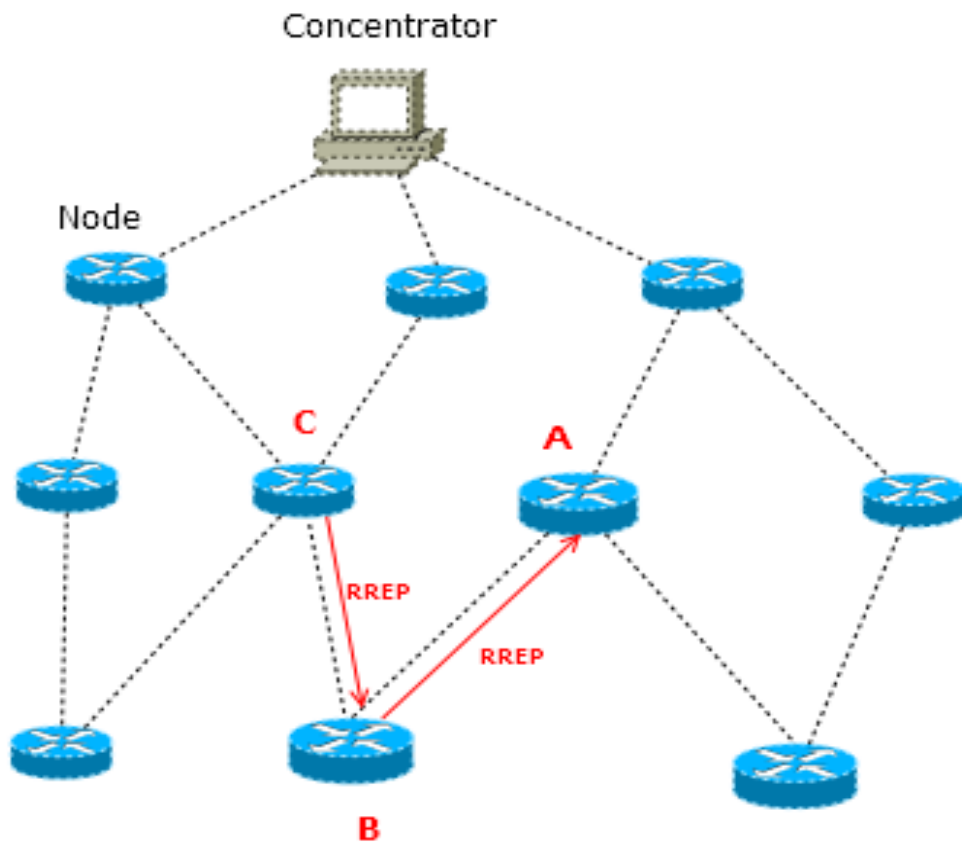

Fig. 2. RREP unicast forwarding in LOADng.

Upon receiving this message, only the terminator can reply by a RREP as specified in Fig. 2 and forward it on unicast, hop-by-hop to the source. When intermediate nodes receive the RREP, they will unicast a proper RREP–ACK to the neighbor from which they received the RREP, in order to notify that the link is bidirectional. If a route is detected broken, an error message can be returned to the source of that data packet. Compared to AODV, intermediate LOADng nodes are not allowed to generate RREPs. As a result, LOADng reduces the size of control messages which is demanded in LLNs. In LOADng there is no more need to the sequence Number included in in AODV messages sent to requesting routers. Also, there is no more Gratuitous RREP; in fact, when an intermediate node has a valid route to the destination the node responds with an RREP on unicast to the source and notify the destination with this Gratuitous message. Thus, message size would be reduced which is definitely suitable to LLNs low-power and memory constraints.

In the other hand, nodes with LOADng protocol do not maintain a precursor list containing the IP address for neighbors likely being a next hop towards each destination – as it is done with AODV protocol –, but they only care about the next hop to forward current packet to its destination. Also, compared to AODV, LOADng Control messages can include TLV (Type-Length-Value) elements, permitting protocol extensions to be developed.

## V. The LOADng-CTP Specification

LOADng Collection Tree Protocol (LOADng-CTP) specified by the authors in [6] is a recent extension to LOADng protocol using a "collection tree" combined with LOADng specification detailed in [4]. The LOADng-CTP extension aims for building bidirectional routes for MP2P and P2MP traffic flow in a sensors network, with low overhead easy maintenance and better performance.

### *A. Protocols Messages*

LOADng-CTP maintains the same packet format of LOADng with some modification on the protocol operations; it introduces two flags to RREQ messages:

- RREQ COLLECTION TREE TRIGGER: This trigger enable routers to discover bidirectional routes in its vicinity.
- RREQ COLLECTION TREE BUILD: When set, permitting the receiving router to build the route to the root.
- HELLO messages: The HELLO messages specified in [11] used for collection tree build in LOADng-CTP protocol are broadcasted by root router and never forwarded by the 1-hop neighbors. These messages are used to identify bidirectional routes.

TABLE I. LOADng-CTP parameters

| Parameters | Description |
|---|---|
| NET TRAVERSAL TIME | It is the maximum time allowed for a packet when traversing the whole network from end to end. |
| RREQ MAX JITTER | It is the maximum jitter for RREQ transmission. |
| HELLO MIN JITTER | It is the minimum jitter for HELLO message transmission. With the following condition HELLO MIN JITTER> 2 × RREQ MAX JITTER |
| HELLOMAXJITTER | It is the maximum jitter for HELLO message transmission. |
| RREPREQUIRED | It is a flag to define if a RREP message is required in order to build routes from the root to sensors while receiving RREQ BUILD message. |

### *B. Router Parameters*

LOADng-CTP uses the router parameters detailed in Table 1 for protocol functioning.

### *C. Protocol Operations*

In this section we describe the basic protocol operation with four algorithms. In the procedure summarized in algorithm 1 all flags are initiated and set by the root router. The collection tree is created by the node requiring to be the root of the collection tree. The root of the collection tree generates RREQ TRIGER set to 1. The originator and destination of the RREQ TRIGGER are set to the address of the root. When a RREQ TRIGGER is generated, a RREQ with RREQ BUILD flag is planned to be sent in 2 × NET TRAVERSAL TIME.

**Algorithm 1: Collection Tree Triggering**

```
If Generate_RREQ (loadngCTP_packet) == true
    and Set (RREQTRIGGER,1)>1 then
<originator>←Route_address
Destination_address← Route_address
End if
while (true)
if (Timer == 2 ×NET TRAVERSAL TIME) then
Generate(RREQBUILD)
Timer←0
End if
    End while
    Return true
    End procedure
```

In the neighbor discovery procedure summarized in Algorithm 2, each router will acquire the list of neighbors with bidirectional (SYM) or unidirectional (HEARD) link and update the routing tables

**Algorithm 2: Bidirectional Neighbour Discovery**

```
If (RREQTRIGGER>0)then
Update(Routing Set, routing tuple<originator>)
Status← HEARD
TriggerReceived← 1
End if
If ((RREQTRIGGER>0) and (TriggerReceived>0)
    then
Jitter← RREQMAXJITTER
Consider_Transmiting_RREQ( loadngCTP_packet )
End if
    if isHelloMessage ( loadngCTP_packet )= true
    then
    if (Routing Tuple==Route_address) then
    Update(Routing Set, routing tuple<originator>)
    Status← SYM
    Discard HelloMessage
    End if
    End if
    return true
    End procedure
```

The algorithm 3 describes the collection tree building process.

When receiving the RREQ BUILD the router validate if the RREQ BUILD was received from a neighbor with a bidirectional link in this case the collection tree is built the routing table and the protocol parameters are updated otherwise the message is discarded.

**Algorithm 3: Collection Tree Building**

```
If((RREQ BUILD>0)and (status== SYM)) then
Build← 1
else if (status== HEARD)
discard RREQ BUILD
End if
End if
If ((RREQ BUILD>0 and (Build>0)) or short path to the
  root then
Insert Routing tuple into Routing Set
R_next_addr ← previous-hop
R_dest_addr ←<originator>
Jitter←  RREQMAXJITTER
Consider_Transmiting_RREQ( loadngCTP_packet )
End if
return true
End procedure
```

The Root-to-Sensor Path Building procedure is detailed in algorithm 4. With the exchange of RREQ TRIGGER and RREQ BUILD messages all the nodes in the collection tree obtained a bidirectional path to the root. In fact, root to sensors path are built by the use of RREP REQUIRED flag set to true and transmitted in unicast to the root and routing tables are updates according to the bidirectional path established.

**Algorithm 4: Root-to-Sensor_Path_Building**

```
If ((RREQ BUILD) >0 and(RREPREQUIRED>0), then
RREP_originator ← Sensor_addr
RREP_destination← R_addr
End if
If (RREPJITTER>0) then
R_dest_addr ← previous-hop
R_next_addr ← previous-hop
R_next_addr ← RREP_destination
End if
return true
End procedure
```

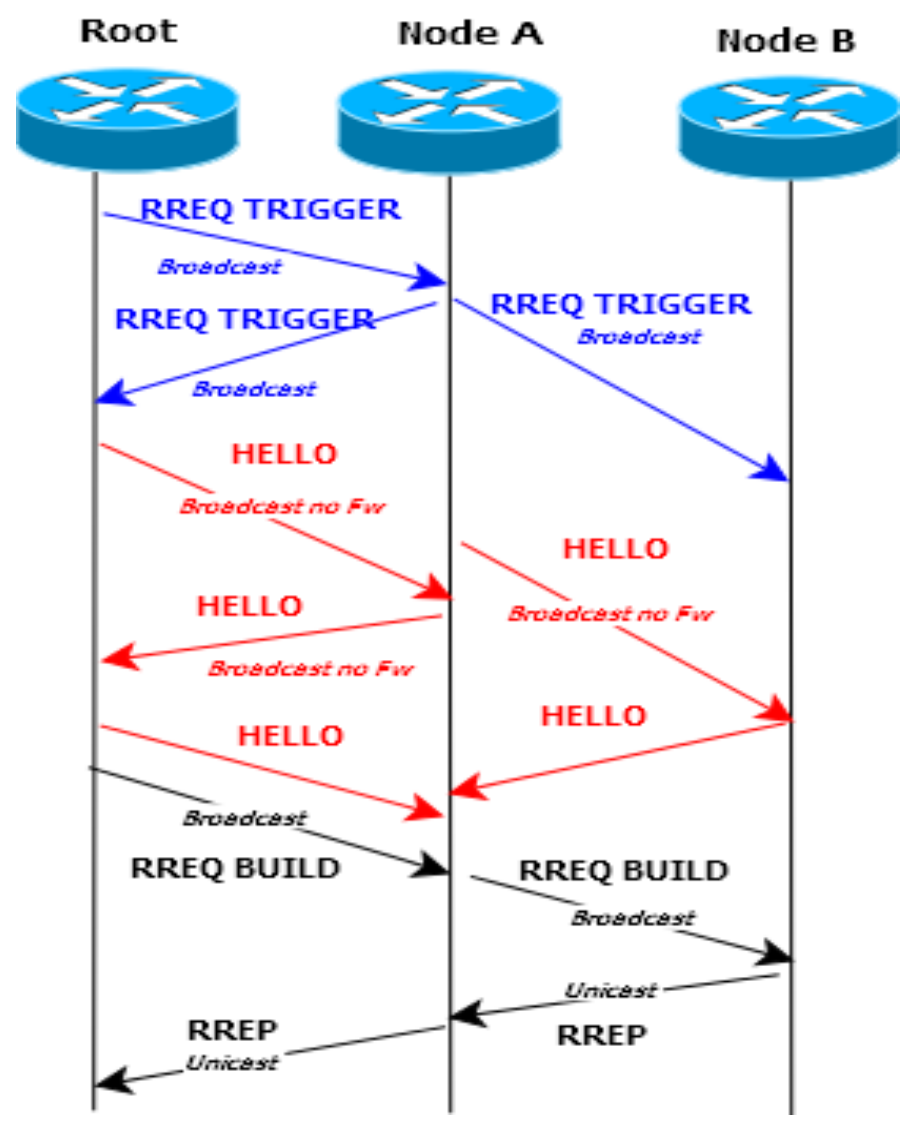

Fig. 3. LOADng-CTP Messages Exchange.

The implementation of the extension LOADng-CTP was successfully integrated into the network layer of Contiki OS with the implementation of the four algorithms described above.

The LOADng-CTP messages exchange functioning is detailed in Fig. 3 with a CTP building between routers A and B resuming the four procedures described in algorithms 1, 2, 3 and 4.

## VI. Performance Evaluation

### *A. Simulation Settings*

In order to understand the performance of the two protocols and LOADng-CTP extension we evaluated routing protocols in terms of packet Delivery ratio (PDR), latency in order to predict how it behave in larger networks, and overhead to describe its power consumption and memory management. The detailed settings of the scenarios studied are detailed in table 2 the values have been averaged over 10 runs;

TABLE II. Cooja simulator parameter setup

| Settings Transport layer UDP | Value |
|---|---|
| Wireless channel model | UDGM Model with Distance Loss |
| Communication range | 250m |
| Distance to the Concentrator | Variable [50-500] Meters |
| Grid Size | 1000*1000 m2 |
| Number of routers | Variable [20/300] |
| Mote type | Tmote Sky |
| Network layer | µIPv6  6LoWPAN |
| MAC layer | CSMA  ContikiMAC |
| Radio interface | CC2420 2.4 GHz IEEE 802.15.4 |
| Simulation time | 8h |

Simulations were completed in a field of 1000 × 1000 meters, with variable amounts of routers positioned randomly , were realistic conditions were approached and different nodes are suggested to interferences. The network scenarios are as described in Table 3 are substance of two different traffic patterns: multipoint-to-point (MP2P), where all routers generate CBR traffic flow by periodic reporting of 512-byte data packet with 60 seconds interval and acknowledgment of each received frame in upward direction of 16 bytes payload, for which the destination always is the sink. And point-to- multipoint (P2MP) traffic with two messages types, acknowledgment of data frames in downward direction every data arrival of 12 bytes payloads and configuration data packet with CBR traffic flow by periodic message of 61-byte payload with 300 seconds interval in downward direction.

TABLE III. Nodes Traffic Pattern

| Node Type | Traffic Pattern |
|---|---|
| Client | MP2P traffic flow by periodic reporting with 60s interval and acknowledgment of each received frame in upward direction. |
| Concentrator | P2MP traffic with two messages types : Acknowledgment of data frames in downward direction every data arrival. Configuration data sent with CBR flow every 300 seconds in downward direction. |

Both LOADng and LOADng-CTP were implemented with C in contiki OS with respect to their authors specification respectively in [10] and [4, 6]. The settings for RPL are listed in Table 4, and for LOADng and LOADng-CTP in Table 5.

TABLE IV. RPL PARAMETERS

| Parameter | Value |
|---|---|
| Mode of operation | non-storing |
| Rank metric | hop count |
| DIOIntervalMin | 2 s |
| DIOIntervalDoublings | 20 |
| DIORedundancyConstant | 1 |
| DAOInterval | 15s |

TABLE V. LOADNG AND LOADNG-CTP PARAMETERS

| Parameter | Value |
|---|---|
| RREQ jitter | 0 - 0.5 s |
| RREQMAXJITTER | 1 s |
| NETTRAVERSALTIME | 10 s |
| HELLO MIN JITTER | 3 s |
| HELLO MAX JITTER | 5 s |
| Route lifetime | 15 s |
| Routing | Mesh routing |

## B. *Simulation Results*

To sum up, findings from the performance evaluation of the three protocols are described in this section both for P2MP and MP2P traffic flows.

- **Point-to-multipoint (P2MP)**

Fig.4 shows the average packet delivery ratios function of variable distance to the concentrator and Fig.5 shows PDR function of variable nodes number, incurring from respectively LOADng-CTP, LOADng and RPL.

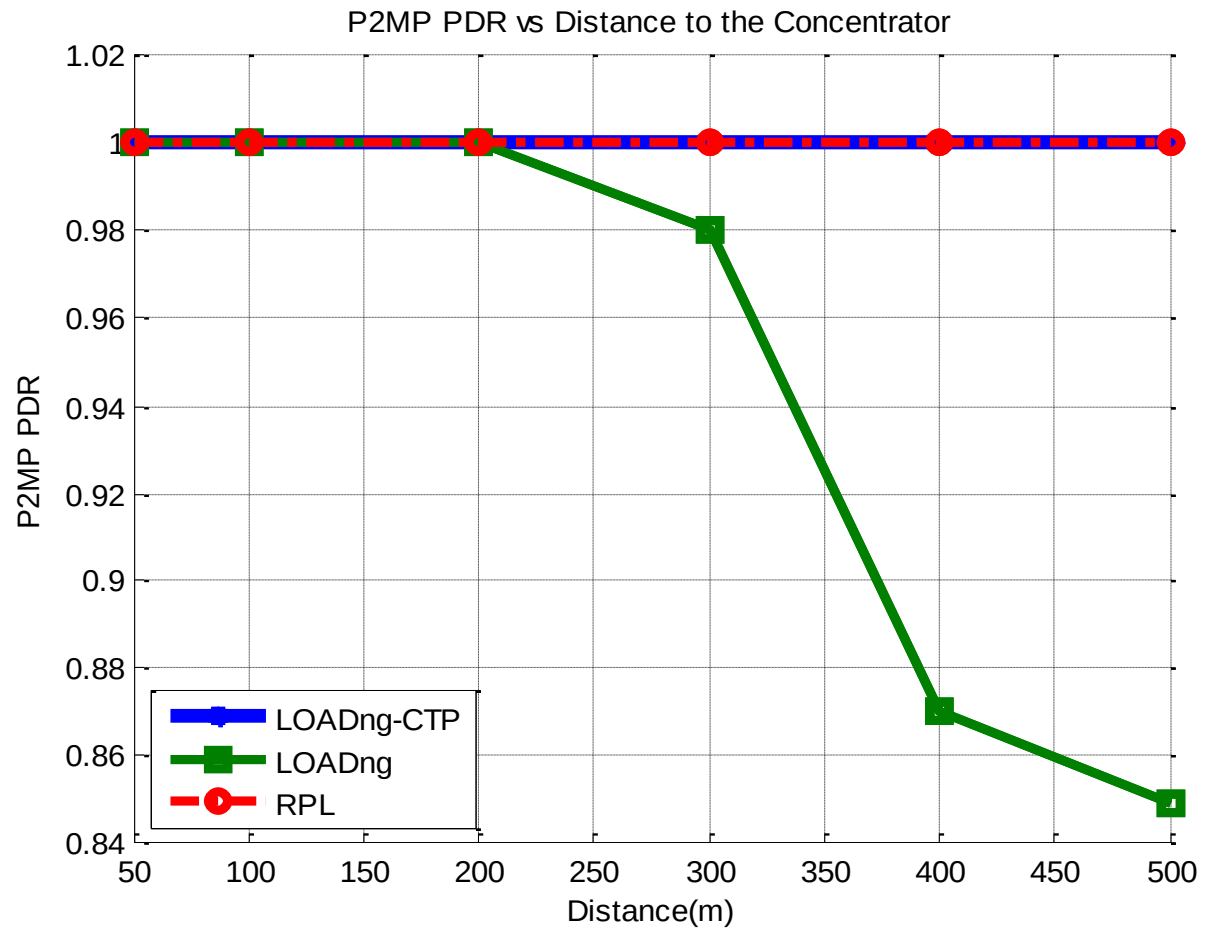

Fig. 4. P2MP PDR function of Distance

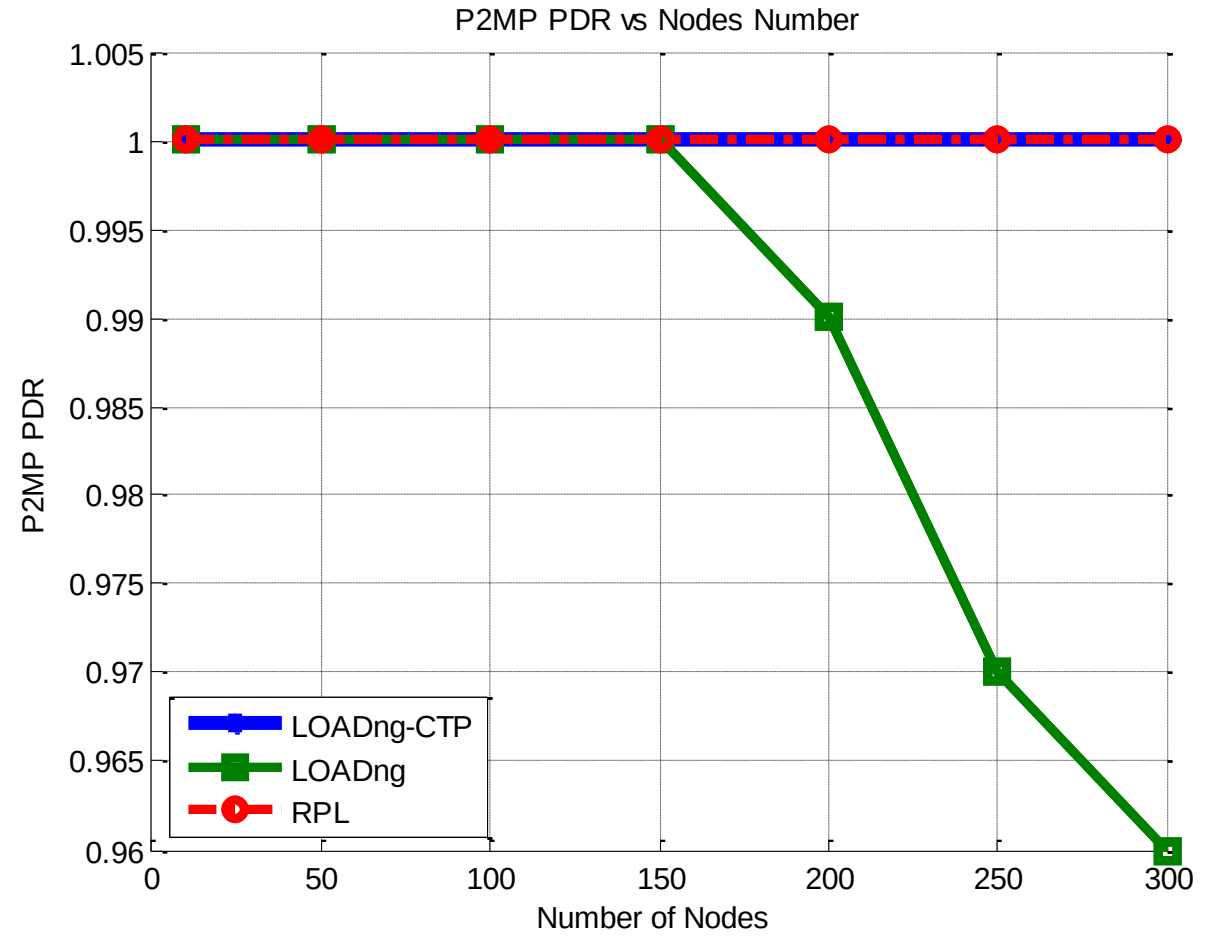

Fig. 5. P2MP PDR function of Nodes Number

As shown by Fig.4 and Fig.5, LOADng packet delivery drops with growing node number to reach 0.98 % for 300 nodes, and 0.96 for 500 meters while PDR remain 100% for RPL and LOADng-CTP.

For downward traffic, Fig. 6 and Fig. 7 shows that LOADng-CTP is very efficient in terms of average latency which is equal to 85 ms compared to RPL 94 ms and LOADng 243 ms when the network is subject to variable distance to the concentrator. While when the network is with increasing number of nodes is variable latency is equal to 81 ms, 95 ms 390 ms for LOADng-CTP, RPL and LOADng respectively.

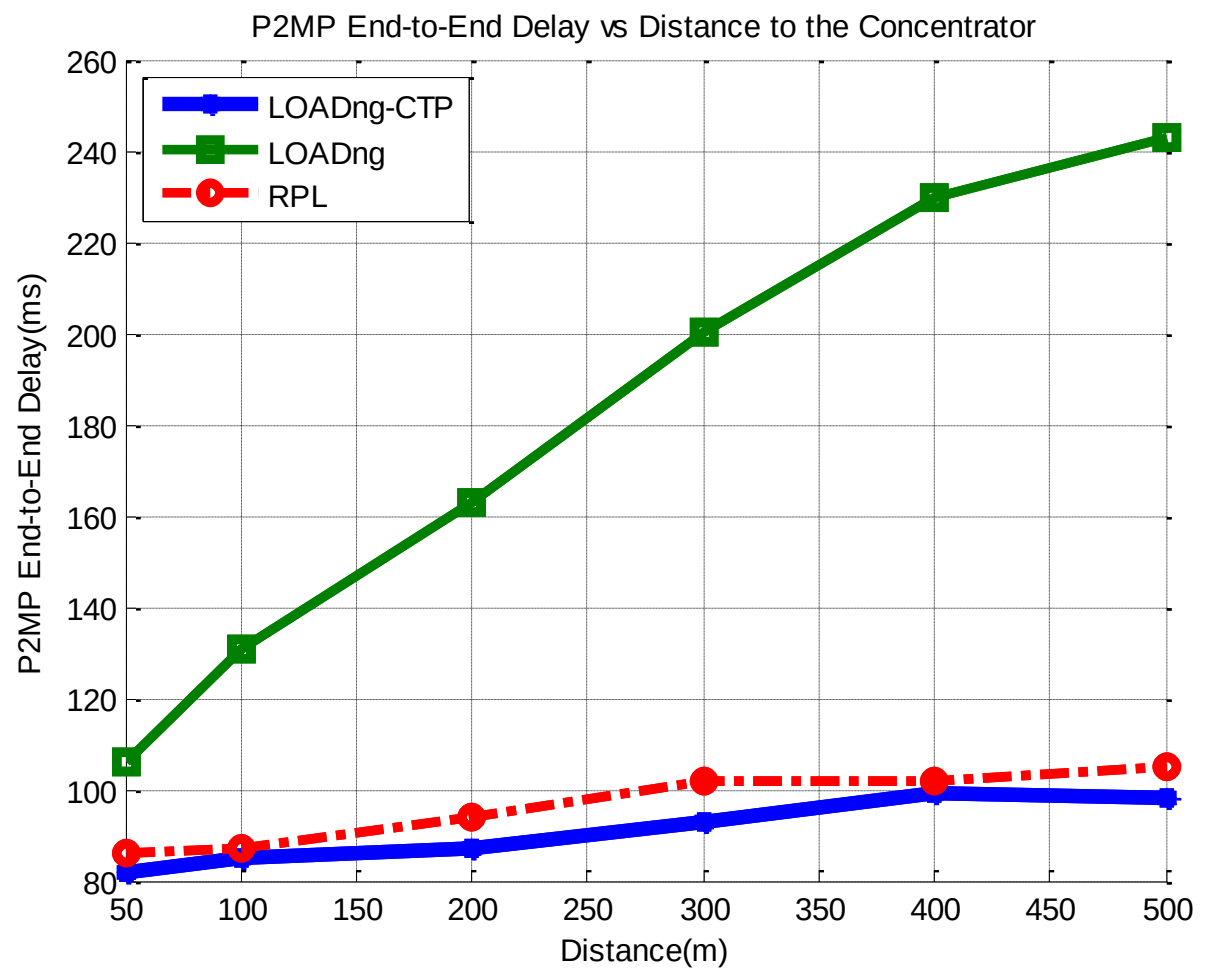

Fig. 6. P2MP End-to-End Delay function of Distance

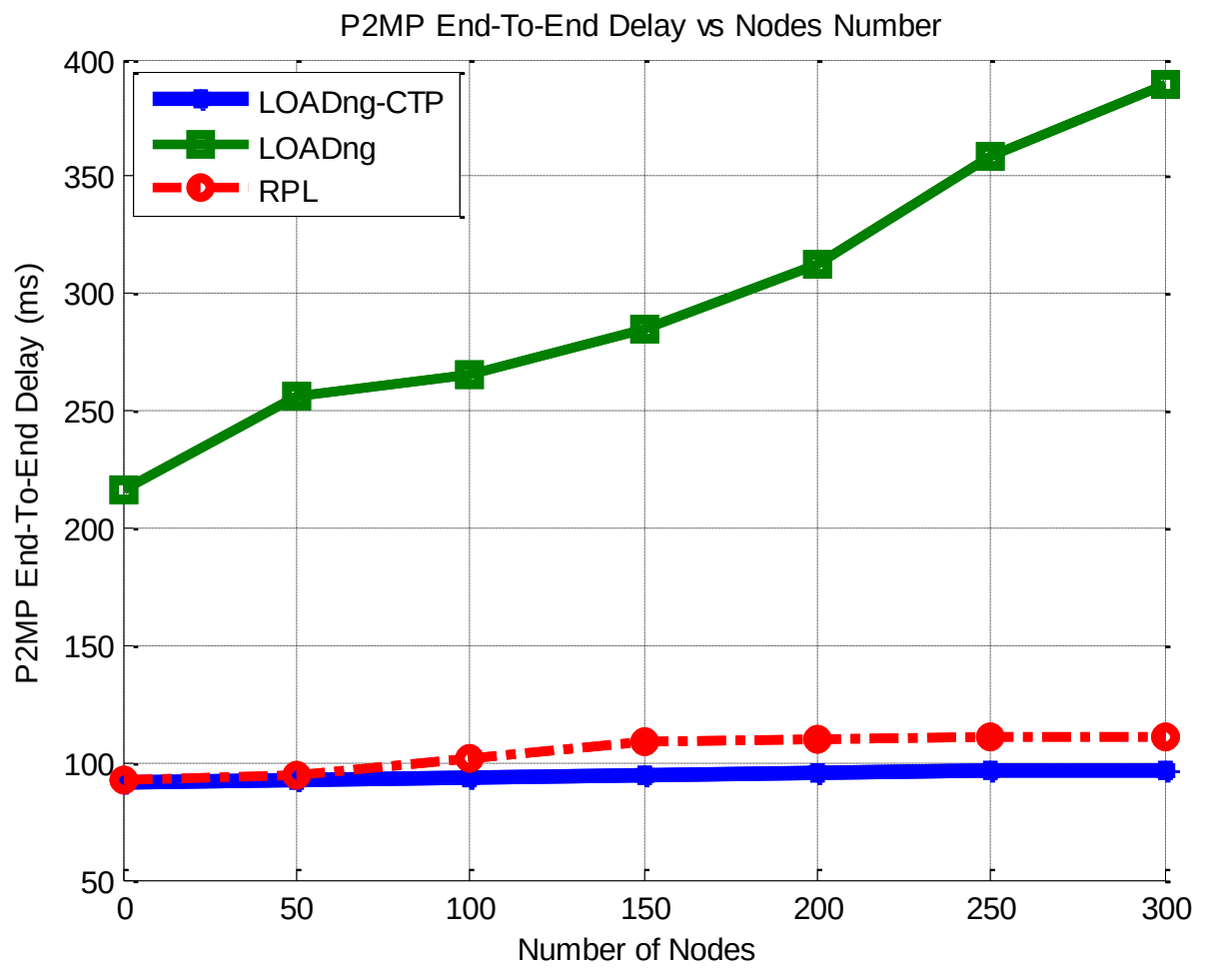

Fig. 7. P2MP End-to-End Delay function of Node Number

- **Multi-to-multipoint (MP2P)**

Fig.8 shows that a packet delivery ratio is 100% for RPL and LOADng-CTP while PDR for LOADng protocol reaches 73% with increasing distance of routers from concentrators.

Whereas, according to Fig.9 we observe significant degradation for the LOADng Protocol delivered packets which reaches 60% when the networks is subject of increasing node number.

For upward traffic, Fig. 10 and Fig. 11 shows that LOADng-CTP is better than LOADng in terms of average latency which is equal to 96 ms compared to RPL 96 ms and LOADng 425 ms when the network is subject to variable distance to the concentrator. Whereas when the network is with increasing number of nodes is variable latency is up to 86 ms, 96 ms and 690 ms for LOADng-CTP, RPL and LOADng respectively.

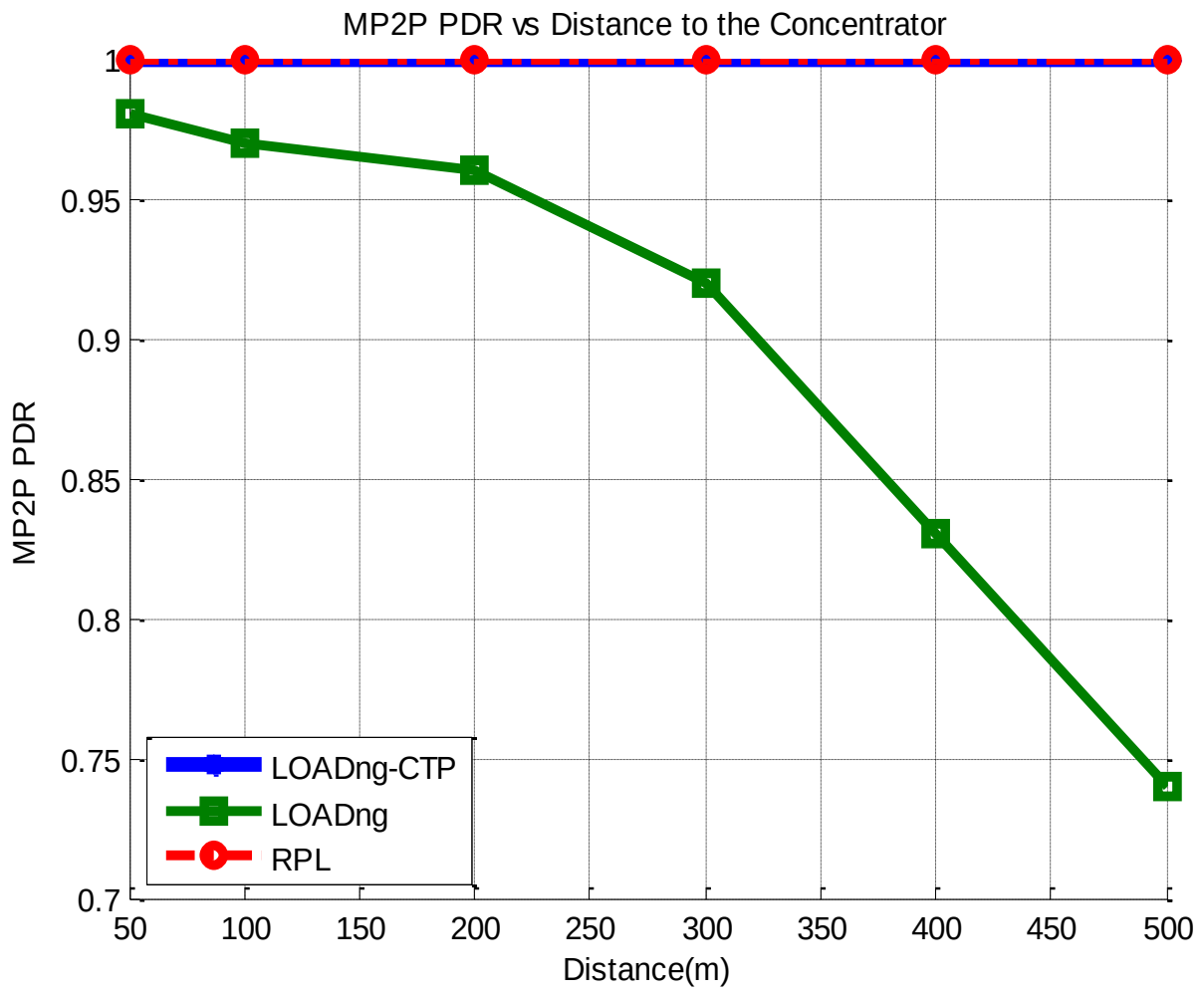

Fig. 8. M2MP PDR function of Distance

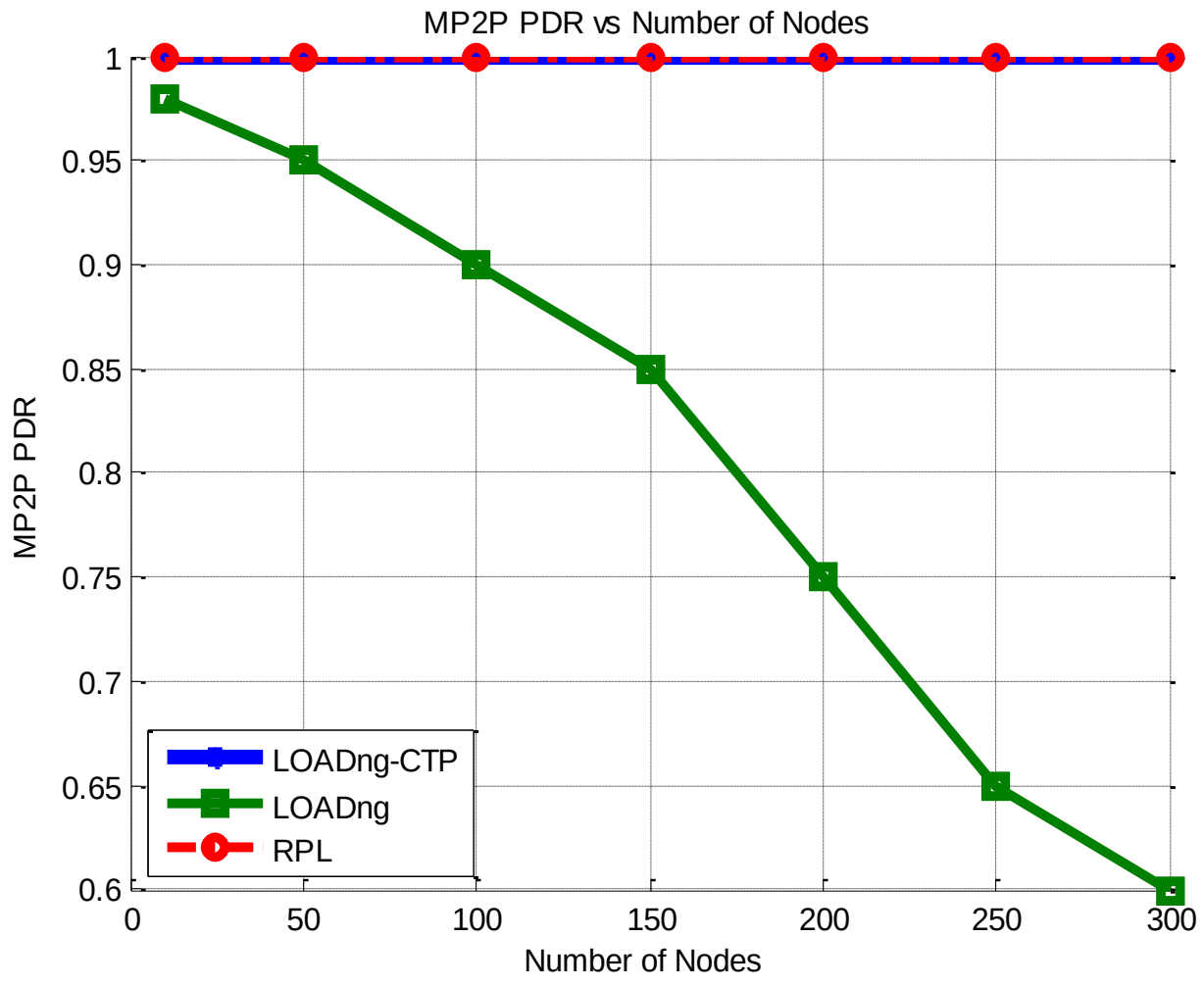

Fig. 9. M2MP PDR function of Nodes Number

- **General Observations**

Fig.12 depicts the Overhead bytes per second in the whole network. The results shows that the overhead for LOADng-CTP has clearly declined compared to LOADng due to collection tree mechanism.

- **Results analysis**

Our results expose the performance of our implementation of LOADng-CTP protocols in Contiki OS. One of its most significant aspects is the considerable reduction of routing overhead due to the smart RREQ used, RREQ flags and the unicast of RREP and its effects in the substantial drop of the messages number and size to maintain routing tables.

Packet delivery ratio for both LOADng-CTP and RPL are 100% with increasing number of nodes whereas PDR drops for LOADng because of MAC layer collision due to LOADng protocol initiation of route discovery for each router.

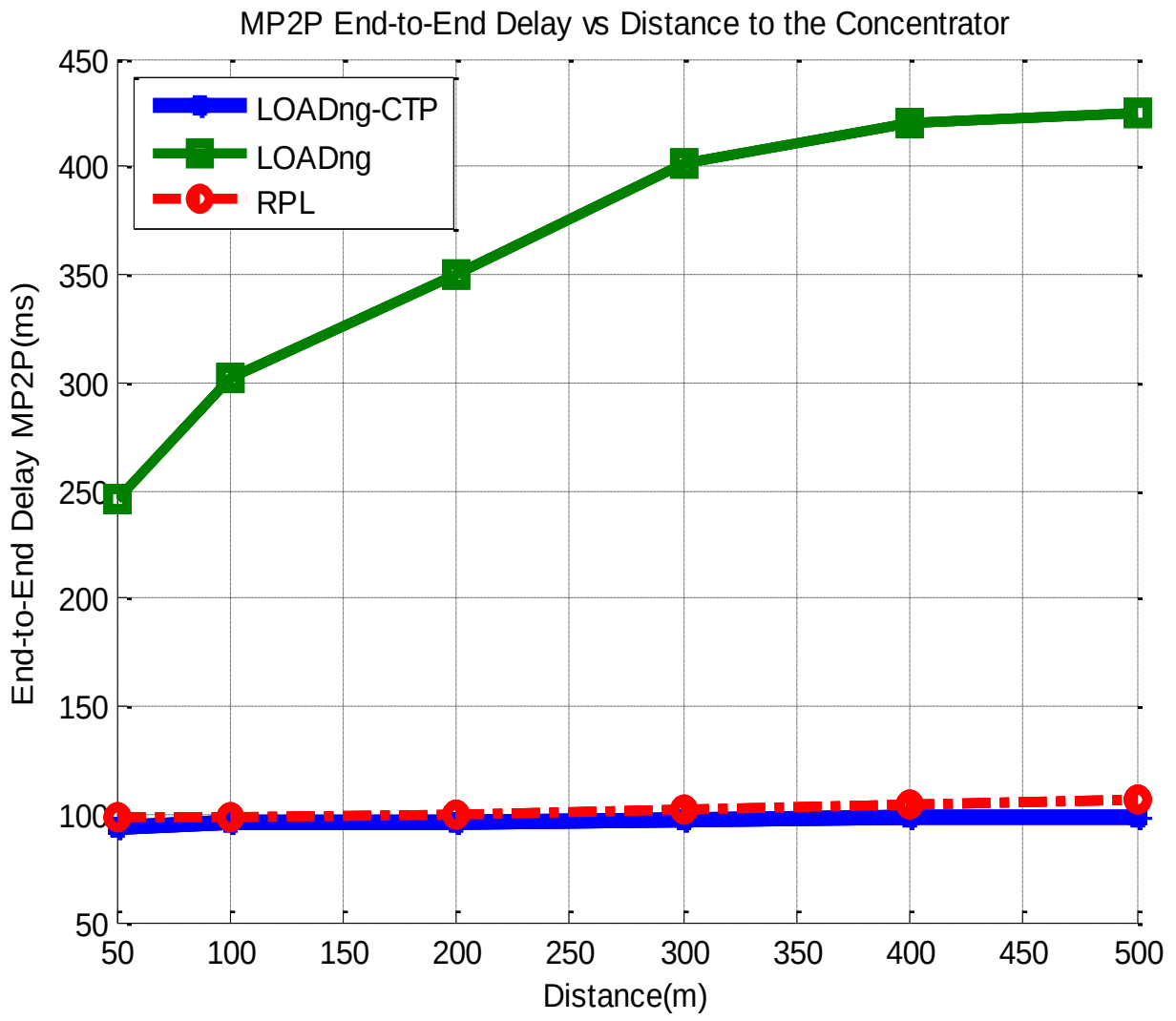

Fig. 10. M2MP End-to-End Delay function of Distance

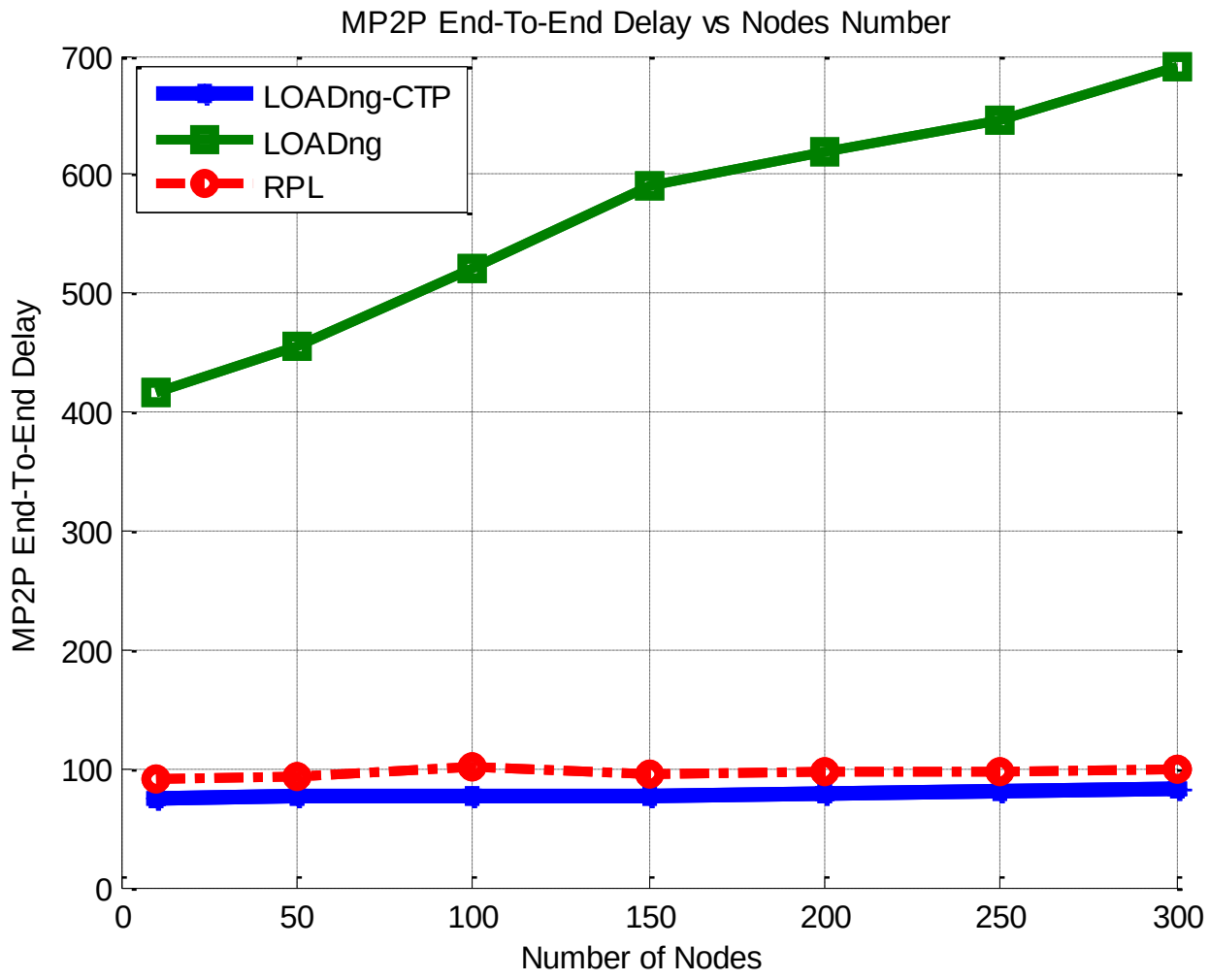

Fig. 11. M2MP End-to-End Delay function of Node Number

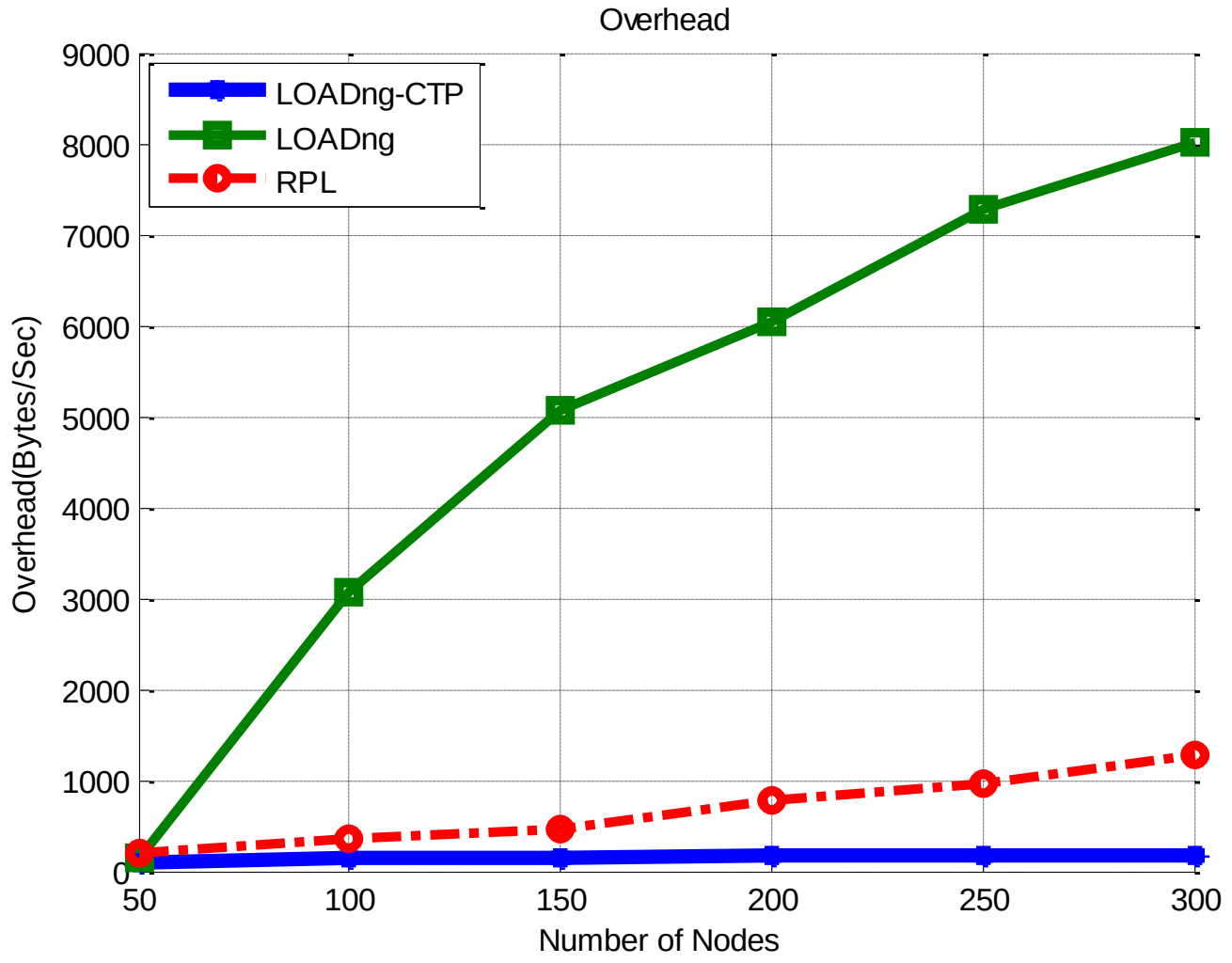

Fig. 12. Overhead bytes per second

LOADng protocols also have bigger End to End delay because route discovery is performed in reactive way in each router while for RPL and LOADng–CTP routes are available created and maintained previously.

## VII. Conclusion

This paper has offered a detailed protocols comparison of LOADng-CTP, to LOADng routing protocol and RPL on behalf of smart metering scenario for MP2P and P2MP traffic types. RPL and LOADng denote two different concepts for routing protocols in LLNs. In spite of its complexity RPL is optimized for specific topologies and traffic patterns. While LOADng represents a part of the ITU-T G.9903 recommendation for smart metering applications. Its strength came from its entirely distributed mode of operation, where paths are discovered on demand and are bidirectional.

The implemented LOADng-CTP extension permit on demand collection trees construction supporting upward traffic from sensors to root in bidirectional traffic scenarios.

Our study reveals that the LOADng-CTP extension harvests better performance than LOADng: higher data delivery ratios, lower delays and lower overhead and LOADng-CTP is comparable to RPL: same data delivery ratios, same delays and lower overhead for bidirectional data traffic in smart metering which make it a better solution than LOADng and RPL for AMI networks.

In our future works we will concentrate on optimizing upward and downward End-to-End Delay for LOADng-CTP protocol to enhance its performance.

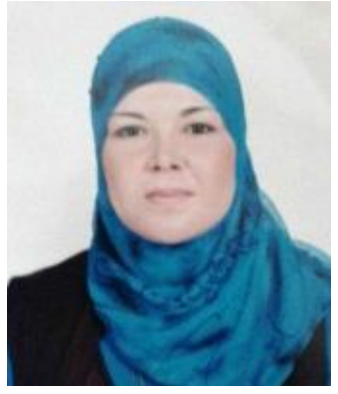

**S. Elyengui**: is a Ph.D. student in the department of communication systems at Tunisian National School of Engineering, University of Tunis El Manar Tunisia. She is a researcher in the area of smart grid communication and networking, SG networks security, AMI applications and M2M communications. She received her Computer Networks Engineer Diploma and a Master degree in new Technologies of Communication and Networking in 2007 and 2011 respectively.

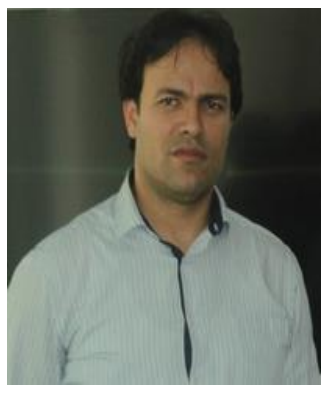

**R. Bouhouchi** received the Master degrees in communication systems from ENIT, the MBA degree s from Mediterranean School of Business, T.E graduated in 2000 from ESPTT, and holds an engineering degree in computer sciences since 2006, as he holds more than eight international certificates in advanced programming and management as ITIL (Information Technologies Infrastructure Library).